\documentclass[aps,prc,twocolumn,groupedaddress,showpacs,floatfix]{revtex4}
\usepackage{graphicx}
\usepackage{dcolumn}
\usepackage{bm}

\newcommand{\rna}{(n,$\alpha$)}

\newcommand{\al}{$\alpha$}

\newcommand{\thcalc}{$T_{1/2,\alpha}^{\rm{calc}}$}
\newcommand{\thexp}{$T_{1/2,\alpha}^{\rm{exp}}$}
\newcommand{\thpre}{$T_{1/2,\alpha}^{\rm{pre}}$}

\begin{document}

{
\title{
$\mathbf\alpha$-nucleus potentials, $\mathbf\alpha$-decay half-lives,
  and shell closures for superheavy nuclei
}

\author{Peter Mohr\footnote{Electronic address: WidmaierMohr@compuserve.de}}
\affiliation{
Diakoniekrankenhaus Schw\"abisch Hall, D-74523 Schw\"abisch Hall,
Germany
}

\date{\today}

\begin{abstract}
Systematic \al -nucleus folding potentials are used to analyze \al
-decay half-lives of superheavy nuclei. Preformation factors of about
several per cent are found for all nuclei under study. The systematic
behavior of the preformation factors and the volume integrals of the
potentials allows to predict \al -decay energies and half-lives for
unknown nuclei. Shell closures can be determined from measured \al
-decay energies using the discontinuity of the volume integral at
shell closures. For the first time a double shell closure is predicted
for $Z_{\rm{magic}} = 132$, $N_{\rm{magic}} = 194$, and
$A_{\rm{magic}} = 326$ from the systematics of folding potentials. The
calculated \al -decay half-lives remain far below one nanosecond for
superheavy nuclei with double shell closure and masses above $A > 300$
independent of the precise knowledge of the magic proton and neutron
numbers.
\end{abstract}

\pacs{23.60.+e,27.90.+b,21.60.Cs,21.60.Gx}

\maketitle
}

%
The \al -decay of superheavy nuclei has been studied intensively in
the last years
\cite{Xu05,Den05,Gam05,Dup05,Don05a,Don05b,Cho05,Dim00,Gar00,Buck93}. In
many papers a simple two-body model was applied \cite{Gur87}, and in
most papers a potential was derived which was able to fit the measured
\al -decay half-lives of the analyzed nuclei. However, most of the
studies (with the exception of \cite{Den05}) did not attempt to use
these potentials for the description of other experimental quantities
like e.g.\ \al\ scattering cross sections or \rna\ or fusion reaction
cross sections.

Therefore, an alternative approach was followed in \cite{Mohr00}. Now
the simple two-body model has been combined with systematic \al
-nucleus folding potentials which are able to describe various
properties, and the ratio between the calculated half-life \thcalc\
and the experimental half-life \thexp\ has been interpreted as
preformation factor $P$ of the \al\ particle in the decaying
nucleus. Besides a systematic behavior of the volume integrals of the
folding potentials, preformation factors of the order of a few per
cent were found for a large number of nuclei \cite{Mohr00,Fuj02}. 

Only for very few light nuclei some levels have been found where a
simple two-body model can exactly reproduce the experimental
half-lives or widths, e.g.\ for $^6$Li = $^2$H $\otimes$ \al\
\cite{Mohr94a} or for $^{8}$Be = \al\ $\otimes $ \al\
\cite{Mohr00,Mohr94b}. Already for $^{20}$Ne = $^{16}$O $\otimes$ \al\
the calculated widths are somewhat larger than the experimentally
observed ones \cite{Mohr05}. Any simple two-body model with a
realistic potential must provide half-lives identical or shorter than
the experimental value, because the two-body model assumes a pure \al\
cluster wave function by definition, whereas any realistic wave
function is given by the sum over many different configurations. Thus,
preformations of a few per cent are a quite natural finding for
superheavy nuclei.

The following ingredients were used in this paper. The \al -nucleus
potential was calculated from a double-folding procedure with an
effective interaction \cite{Sat79,Kob84,Mohr00}. The nuclear densities
were taken from \cite{Vri87} for the \al\ particle and derived from
the two-parameter Fermi distributions for $^{232}$Th and $^{238}$U in
\cite{Vri87} with properly scaled radius parameter $r \sim
A_T^{1/3}$. The total potential is given by the sum of the nuclear
potential $V_N(r)$ and the Coulomb potential $V_C(r)$:
\begin{equation}
V(r) = V_N(r) + V_C(r) = \lambda \, V_F(r) + V_C(r)
\label{eq:vtot}
\end{equation}
The Coulomb potential is taken in the usual form of a homogeneously
charged sphere where the Coulomb radius $R_C$ has been chosen
identically with the $rms$ radius of the folding potential $V_F$, and
the folding potential $V_F$ is scaled by a strength parameter
$\lambda$ which is of the order of $1.0 - 1.3$. This leads to volume
integrals of about $J_R \approx 300$\,MeV\,fm$^3$ for all nuclei under
study and is in agreement with systematic \al -nucleus potentials
derived from elastic scattering
\cite{Atz96,Mohr97,Ful01,Gal05,Dem02,Avr03,Avr05}. (Note that as usual
the negative sign of $J_R$ is omitted in this work.) Bound state
properties of $^{212}$Po = $^{208}$Pb $\otimes$ \al\ have been
analyzed successfully using the same potential \cite{Hoy94}. The
centrifugal potential has been omitted for $L = 0$ decays. The
following study is restricted to even-even nuclei because the
additional centrifugal barrier may influence the \al -decay half-life
for decays with $L \ne 0$.

The quotations of the volume integral $J_R$ and the potential strength
parameter $\lambda$ are practically equivalent for this paper. If one
wants to compare this work to folding potentials with a different
nucleon-nucleon interaction or even to potentials with a different
parametrization (e.g., Woods-Saxon potentials), the volume integral
$J_R$ has to be used. Therefore the following discussion is restricted
to volume integrals. Nevertheless, most figures provide both quantities
$J_R$ and $\lambda$. 

The $\alpha$-decay width $\Gamma_\alpha$ is given by the following
formulae \cite{Gur87}:
\begin{equation}
\Gamma_\alpha = P F \frac{\hbar^2}{4\mu} 
\exp{\left[ -2 \int_{r_2}^{r_3} k(r) dr \right]}
\label{eq:gamma}
\end{equation}
with the preformation factor $P$, the normalization factor $F$ 
\begin{equation}
F \int_{r_1}^{r_2} \frac{dr}{k(r)} = 1
\label{eq:f}
\end{equation}
and the wave number $k(r)$
\begin{equation}
k(r) = \sqrt{ \frac{2\mu}{\hbar^2}\left|E - V(r)\right|} \quad \quad .
\label{eq:k}
\end{equation}
$\mu$ is the reduced mass and $E$ is the decay energy of the
$\alpha$-decay which was taken from the computer files based on the
mass table of Ref.~\cite{Audi03} or from Table 1 of \cite{Xu05}. The
$r_i$ are the classical turning points. For $0^+ \rightarrow 0^+$
$s$-wave decay the inner turning point is at $r_1 = 0$. $r_2$ varies
around 9\,fm, and $r_3$ varies strongly depending on the energy. The
decay width $\Gamma_\alpha$ is related to the half-life by the
well-known relation $\Gamma_\alpha = \hbar \ln{2} /
T_{{1/2},\alpha}$. Following Eq.~(\ref{eq:gamma}), the preformation
factor may also be obtained as
\begin{equation}
P = \frac{ T_{1/2,\alpha}^{\rm{calc}} }{ T_{1/2,\alpha}^{\rm{exp}} }
\label{eq:pre}
\end{equation}
where $\Gamma_\alpha$ or \thcalc\ are calculated from
Eq.~(\ref{eq:gamma}) with $P = 1$. For completeness, I define the here
predicted half-life for unknown nuclei as \thpre $ = $\thcalc $/P$.

The potential strength parameter $\lambda$ was adjusted to the
energy of the $\alpha$ particle in the $\alpha$ emitter
$(A+4) = A \otimes \alpha$. The number of nodes of the bound state
wave function was taken from the Wildermuth condition
\begin{equation}
Q = 2N + L = \sum_{i=1}^4 (2n_i + l_i) = \sum_{i=1}^4 q_i
\label{eq:wild}
\end{equation}
where $Q$ is the number of oscillator quanta,
$N$ is the number of nodes and $L$ the relative angular
momentum of the $\alpha$-core wave function, and
$q_i = 2n_i + l_i$ are the corresponding quantum numbers
of the nucleons in the $\alpha$ cluster. I have taken
$q = 5$ for $82 < Z,N \le 126$ and
$q = 6$ for $N > 126$
where $Z$ and $N$ are the proton and neutron number of the daughter
nucleus. The above definition of $Q$ slightly deviates from the
semi-classical Bohr-Sommerfeld quantum number $G$ as mostly used. One
finds $G \approx 22.5$ for all nuclei with $Q = 22$.

Various attempts have been made to determine the preformation factors
$P$ experimentally or theoretically \cite{Ton79,Iri86,Var92}. The
usage of a simple two-body wave function in connection with the
Wildermuth condition Eq.~(\ref{eq:wild}) is obviously a quite simple
approximation for the description of the complex many-body wave
function of a superheavy nucleus which was analyzed e.g.\ in
\cite{Flo00,Del02,Afa03,Del04,Zha05}. Nevertheless, the preformation
factor defined as ratio $P = $ \thcalc /\thexp \ in Eq.~(\ref{eq:pre})
may be understood as effective preformation factor. The obtained
values for $P$ do only show small variations and can thus be used for
the prediction of half-lives of unknown superheavy nuclei in a
consistent way. A full discussion of preformation factors is beyond
the scope of the present paper.

\begin{figure}
\includegraphics[ bb = 55 60 490 355, width = 70 mm, clip]{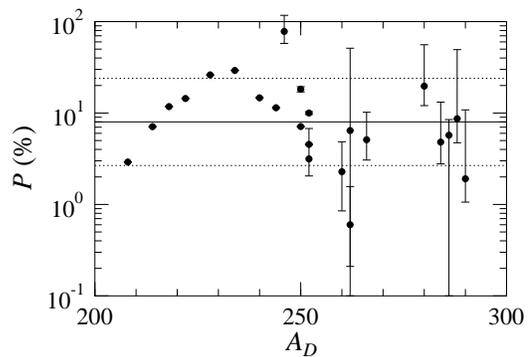}
\caption{
  \label{fig:pre} 
  The preformation factors $P$ are shown for several superheavy
  $\alpha$ emitters. The horizontal lines indicate an average value of
  $P \approx 8$\,\% (full line) and typical uncertainties of a factor
  of three (dotted lines).
}
\end{figure}

The resulting preformation factors $P$ for even-even nuclei are shown
in Fig.~\ref{fig:pre}. An average value of $P \approx 8$\,\% is
found. Almost all results lie within a bar of uncertainty of a factor
of three. This uncertainty is identical to the results of
\cite{Xu05,Den05}. However, the values for $P$ are much smaller in
this work (see discussion above). A table of the results will be given
in a forthcoming paper.

There are two different ways in this simple two-body model to obtain
larger \al -decay half-lives \thcalc\ and thus larger preformation
factors $P$ as derived from Eq.~(\ref{eq:pre}). First, very narrow
potentials can be used. In this case the attractive nuclear potential
becomes negligible in the region of the Coulomb barrier thus
effectively increasing the barrier and increasing the \al -decay
half-life. This idea was followed e.g.\ in \cite{Buck93}, and the
differences to the systematic folding potential in the present work
are illustrated in Fig.~1 of \cite{Mohr00}. A very narrow potential as
used in \cite{Buck93} is probably not able to describe quantities
beside the \al -decay half-life. Second, a smaller quantum number ($G$
or $Q$) can be used. This idea was followed in \cite{Xu05}. In this
case the attractive nuclear potential is reduced at all radii, thus
again effectively increasing the Coulomb barrier and the \al -decay
half-life. Many quantities are mainly sensitive to the tail of the
wave functions at large radii outside the nuclear potential which
leads to discrete ambiguities for the volume integral $J_R$ of \al
-nucleus potentials (the so-called ``family problem''). However, it
has been found in the last years that systematic \al -nucleus folding
potentials have volume integrals $J_R$ around 300\,MeV\,fm$^3$
\cite{Atz96,Mohr97,Ful01,Gal05,Dem02,Avr03,Avr05} compatible with the
quantum number $Q$ used in the present work and incompatible with the
smaller $G$ used in \cite{Xu05}.

\begin{figure}
\includegraphics[ bb = 55 60 490 265, width = 70 mm, clip]{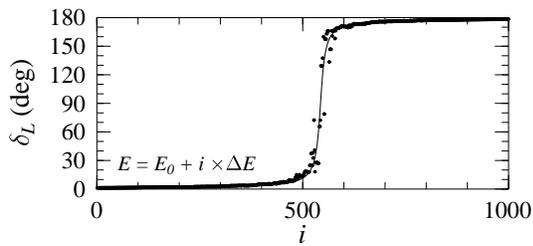}
\caption{
  \label{fig:phase} 
  Phase shift $\delta_L$ for the $L = 0$ partial wave for the system
  $^{212}$Po = $^{208}$Pb $\otimes$ \al . The derived width from
  Eq.~(\ref{eq:phase}) is $\Gamma = 3.52 \times 10^{-14}$\,MeV. Note
  the extremely small stepsize of the calculation of $\Delta E = 2.0
  \times 10^{-15}$\,MeV!
}
\end{figure}

In principle, the application of a semi-classical model is not
necessary for the calculation of \al -decay half-lives or widths. From
the potential in Eq.~(\ref{eq:vtot}) one can directly calculate the
wave function and the width of the decaying state. However, in
practice this is difficult because of the low energies and extremely
small widths of the states. For $^{212}$Po = $^{208}$Pb $\otimes$ \al\
such a fully quantum-mechanical calculation is possible at the limits
of numerical stability. Fig.~\ref{fig:phase} shows the scattering
phase shift $\delta_L$ for the $L = 0$ partial wave as a function of
energy which is given by $E = E_0 + i \times \Delta E$ with $E_0 =
8.954088523002$\,MeV and $\Delta E = 2 \times 10^{-15}$\,MeV. The
points are the results of a phase shift calculation, the line is a fit
to the points using the formula for narrow resonances
\begin{equation}
\delta_L(E) = \arctan{\frac{\Gamma}{2(E_R-E)}}
\label{eq:phase}
\end{equation}
with $E_R = 8.9541$\,MeV and $\Gamma = 3.52 \times 10^{-8}$\,eV which
translates to \thcalc\  $= 13.0$\,ns. The semi-classical approximation
yields \thcalc\ $= 8.7$\,ns which is about 30\,\% lower than the
precise quantum-mechanical value. A similar result is obtained for the
decay of $^{8}$Be where one finds $\Gamma_\alpha = 6.7$\,eV for the
quantum-mechanical calculation and $\Gamma_\alpha = 10.5$\,eV for the
semi-classical approximation. As pointed out above, the preformation
factors are close to unity for $^8$Be with $P = 100$\,\% (65\,\%) for the
quantum-mechanical (semi-classical) calculation and of the order of a
few per cent for $^{212}$Po with $P = 4.3$\,\% (2.9\,\%). These
results confirm the applicability of the semi-classical model within
uncertainties of about 30\,\%.

\begin{figure}
\includegraphics[ bb = 50 60 490 677, width = 70 mm, clip]{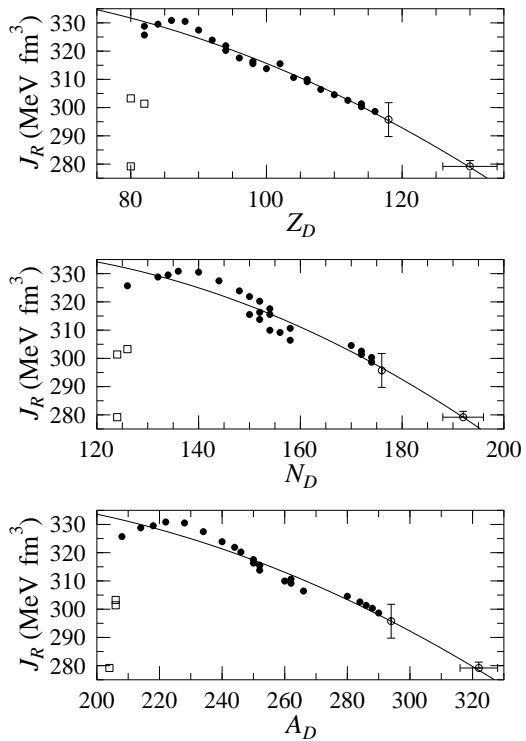}
\caption{
  \label{fig:jr} 
  Volume integrals $J_R$ for superheavy nuclei as a function of $Z_D$
  (upper), $N_D$ (middle), and $A_D$ (lower). Within a major shell one
  finds a smooth decrease of $J_R$ with a minimum value around $J_R
  \approx 280$\,MeV\,fm$^3$. From the fits to the data points one can
  directly see that $J_R$ drops below 280\,MeV\,fm$^3$ at $Z \approx
  130$, $N \approx 192$, and $A \approx 322$ leading to the magic
  numbers $Z_{\rm{magic}} = 132$, $N_{\rm{magic}} = 194$, and
  $A_{\rm{magic}} = 326$. The open circles are extrapolations for
  $^{294}$118 and the lower limit of $J_R > 279$\,MeV\,fm$^3$ close to
  the next closed shells (see text). The open squares are nuclei with
  $Q = 18$ ($^{208}$Pb = $^{204}$Hg $\otimes$ \al ) and $Q = 20$
  ($^{210}$Pb = $^{206}$Hg $\otimes$ \al , $^{210}$Po = $^{206}$Pb
  $\otimes$ \al ).  
}
\end{figure}

It is interesting to use the systematic folding potentials for the
prediction of properties of unknown superheavy nuclei like \al -decay
energies, \al -decay half-lives, and shell closures above $N,Z =
126$. The basic building block is the smooth behavior of the strength
parameter $\lambda$ of the folding potential and the resulting volume
integrals $J_R$ (see Fig.~\ref{fig:jr} and Table I of \cite{Mohr00}).

Within one major shell, one finds variations of $J_R$ from about
$J_R \approx 335$\,MeV\,fm$^3$ at the lower end of a shell to about
$J_R \approx 280$\,MeV\,fm$^3$ at the upper end of a shell. Between
neighboring nuclei the variation in $J_R$ is below $\Delta J_R <
5$\,MeV\,fm$^3$. This allows first the determination of \al -decay
energies for up to now unknown nuclei. As an example, one finds for
the decay of $^{298}$120 $\rightarrow$ $^{294}$118 a volume integral of
$J_R \approx 296$\,MeV\,fm$^3$ corresponding $\lambda = 1.138$. This
leads to a decay energy of $E = 12.87$\,MeV. The \al -decay half-life
can be estimated using the given energy and an average preformation
factor of $P \approx 8$\,\% leading to \thpre $\approx
8\,\mu$s. Whereas the uncertainties for the volume integral $J_R$ and
the derived \al -decay energy are small, the uncertainty of the \al
-decay half-life is strong because of the exponential energy
dependence. For a potential strength enhanced (reduced) by 2\,\% one
finds the \al -decay energy $E = 10.70$\,MeV ($E = 14.98$\,MeV) and
\thpre $ = 0.97$\,s (\thpre $ = 1.6$\,ns) again using $P = 8$\,\%. A
variation of the potential strength of 1\,\% corresponds to a
variation of the \al -decay energy of about 1\,MeV which is
comparable to the uncertainties of mass formulae \cite{Sto05}. As
usual, the reliability of such an extrapolation decreases for nuclei
with masses far above the heaviest known nuclei. However, the
uncertainties for closed-shell nuclei remain small because of the
well-defined volume integral $J_R$ for such nuclei which can be
studied at the shelle closures at $N = 82$, $Z = 82$, and $N = 126$.

\begin{figure}
\includegraphics[ bb = 50 60 490 410, width = 70 mm, clip]{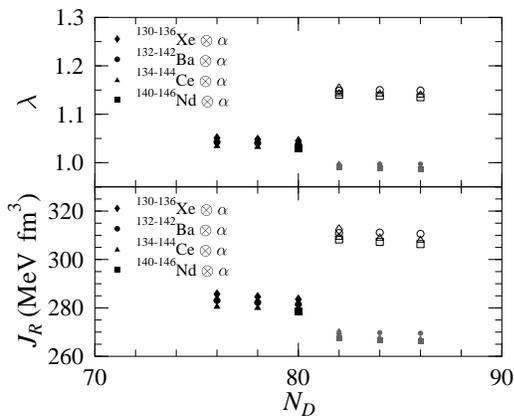}
\caption{
  \label{fig:n82} 
  Potential strength parameter $\lambda$ (upper part) and volume
  integrals $J_R$ (lower) around the shell closure $N = 82$ for
  $^{130-136}$Xe $\otimes$ $\alpha$ (diamonds), $^{132-142}$Ba
  $\otimes$ $\alpha$ (circles), $^{134-144}$Ce $\otimes$ $\alpha$
  (triangles), and $^{140-146}$Nd $\otimes$ $\alpha$ (squares)
  isotopes. One can clearly see the discontinuity at the shell closure
  $N = 82$ (see text).  
}
\end{figure}

Shell closures can be seen as discontinuities in the volume integrals,
see Figs.~\ref{fig:jr} and \ref{fig:n82}. Whereas the variation
between neighboring nuclei remains below $\Delta J_R <
5$\,MeV\,fm$^3$, at shell closures one finds a strong increase of
$J_R$ which is directly related to the increase of the quantum number
$Q$. Because shell closures are not known a priori for superheavy
nuclei, Fig.~\ref{fig:n82} analyzes the volume integrals around the
shell closure at $N = 82$ for Xe, Ba, Ce and Nd isotopes. Below $N =
82$, the wave functions are characterized by $Q = 16$ (full black
symbols), and the volume integrals are slightly above $J_R =
280$\,MeV\,fm$^3$. Above $N = 82$ one finds volume integrals around
$310$\,MeV\,fm$^3$ for wave functions with $Q = 18$ (open symbols; see
also \cite{Mohr00}). The small grey symbols are calculated above the
shell closure at $N = 82$ without increase of the quantum number
$Q$. Here one finds low volume integrals significantly below $J_R =
280$\,MeV\,fm$^3$ which differ by more than $\Delta J_R =
10$\,MeV\,fm$^3$ from the neighboring values. The behavior of the
potential strength parameter $\lambda$ is similar to $J_R$ (see
Fig.~\ref{fig:n82}).

A similar behavior is found at the shell closures at $Z = 82$ and $N =
126$ around $^{208}$Pb. Compared to neighboring values, the volume
integrals $J_R$ are reduced by more than $\Delta J_R =
10$\,MeV\,fm$^3$ if one neglects the increase of the quantum number
$Q$, and the absolute values of the volume integrals drop below $J_R =
280$\,MeV\,fm$^3$. Consequently, changes in $J_R$ by about
10\,MeV\,fm$^3$ or values of $J_R$ significantly below
280\,MeV\,fm$^3$ are clear indications for the crossing of a major
shell. Because the determination of the volume integral $J_R$ requires
only the knowledge of the \al -decay energy, the measurement of one
single quantity may be sufficient for the determination of a double
closure: as soon as $J_R$ drops below 280\,MeV\,fm$^3$, magic neutron
or proton numbers have been crossed!

This systematic behavior of \al -nucleus potentials allows further a
prediction of magic numbers in a yet unknown mass region above $A >
300$. The smooth energy dependence of the volume integrals $J_R$ in
Fig.~\ref{fig:jr} is fitted using a second-order polynomial for all
nuclei with $Q = 22$ (full lines in Fig.~\ref{fig:jr}). $J_R$ drops
below 280\,MeV\,fm$^3$ at $Z = 130$, $N = 192$, and $A = 322$ which
means that the nucleus $^{326}$132 = $^{322}$130 $\otimes$ \al\ is the
heaviest nucleus which can be described using a potential with $J_R >
280$\,MeV\,fm$^3$ and $Q = 22$. Increasing $Z$ or $N$ leads to $J_R$
below its lower limit, and thus the magic numbers $Z_{\rm{magic}} =
132 \pm 4$, $N_{\rm{magic}} = 194 \pm 4$, and $A_{\rm{magic}} = 326
\pm 6$ can be derived from Fig.~\ref{fig:jr}.

The \al -decay half-life of the doubly-magic nucleus with
$Z_{\rm{magic}} = 132$, $N_{\rm{magic}} = 194$, and $A_{\rm{magic}} =
326$ can be calculated using the volume integral $J_R =
279.2$\,MeV\,fm$^3$ (taken from $^{208}$Pb = $^{204}$Hg $\otimes$ \al
). One finds the energy $E = 18.26$\,MeV and the corresponding
half-life \thcalc\ $= 1.16 \times 10^{-12}$\,s with $P = 1$. Again
using $P = 8$\,\%, a realistic prediction of the half-life is \thpre\
= $1.5 \times 10^{-11}$\,s. Including the uncertainty of $P$, the
half-life remains below $10^{-10}$\,s. The uncertainty of the volume
integral $J_R$ at closed shells is smaller than 1\,\%. Increasing
$J_R$ by 1\,\% reduces the \al -decay energy by about 1\,MeV and
increases the \al -decay half-life by about a factor of 20. In any
case, the half-life remains below 1\,ns.

The lower limit of $J_R$ has also been applied for the
calculation of the half-life of $^{310}$126 $\rightarrow$ $^{306}$124
with the widely discussed shell closures at $Z = 126$ and $N = 184$
(e.g., \cite{Lir02,Bar05}); but also other magic numbers have been
discussed (e.g.\ \cite{Sil04}). Here one obtains the \al -decay energy
$E = 18.82$\,MeV and the corresponding \al -decay half-life \thcalc\ $=
2.1 \times 10^{-14}$\,s ($P = 1$). The realistic prediction using $P =
8$\,\% is \thpre\ $= 2.6 \times 10^{-13}$\,s, and including all
uncertainties the half-life remains far below $10^{-11}$\,s. The above
calculations indicate clearly that one cannot expect that any
superheavy nucleus above $A > 300$ with magic proton and neutron
numbers (whatever these numbers are) has a half-life significantly
above one nanosecond.

\begin{figure}
\includegraphics[ bb = 50 60 490 410, width = 70 mm, clip]{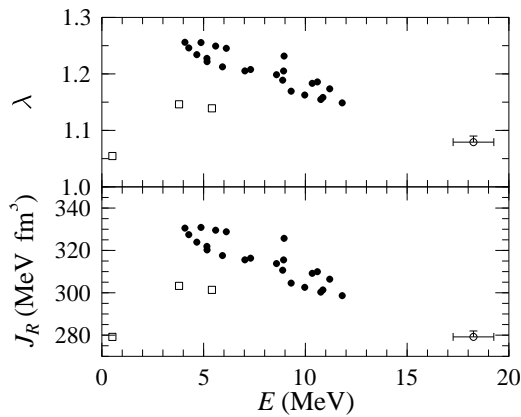}
\caption{
  \label{fig:qlam} 
  Potential strength parameter $\lambda$ (upper part) and volume
  integrals $J_R$ (lower) versus decay energy $E$ for superheavy
  nuclei. Known nuclei are shown as full circles. The extrapolated
  doubly-magic nucleus with $Z_{\rm{magic}} = 132$, $N_{\rm{magic}} =
  194$, and $A_{\rm{magic}} = 326$ is shown as open circle. The open
  squares are nuclei with $Q = 18$ and $Q = 20$ (see also
  Fig.~\ref{fig:jr}). 
}
\end{figure}

\begin{figure}
\includegraphics[ bb = 50 60 490 410, width = 70 mm, clip]{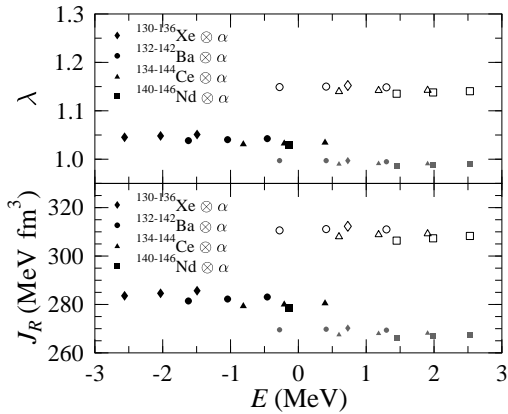}
\caption{
  \label{fig:n82qlam} 
  Potential strength parameter $\lambda$ (upper part) and volume
  integrals $J_R$ (lower) versus decay energy $E$ around the shell
  closure $N = 82$ for $^{130-136}$Xe $\otimes$ $\alpha$ (diamonds),
  $^{132-142}$Ba $\otimes$ $\alpha$ (circles), $^{134-144}$Ce
  $\otimes$ $\alpha$ (triangles), and $^{140-146}$Nd $\otimes$
  $\alpha$ (squares) isotopes (see text; description of symbols see
  Fig.~\ref{fig:n82}).
}
\end{figure}

Because of the significant variation of the decay energy $E$ from
about 4\,MeV to about 12\,MeV for known superheavy nuclei and up to
about 20\,MeV for the predicted but yet unknown doubly-magic
superheavy nuclei one may expect a correlation between the potential
strength parameter $\lambda$ or the volume integral $J_R$ and the
decay energy $E$. This relation is analyzed in Fig.~\ref{fig:qlam} for
superheavy nuclei and Fig.~\ref{fig:n82qlam} for nuclei around $N = 82$.

Fig.~\ref{fig:qlam} seems to indicate that larger decay energies $E$
are correlated to smaller volume integrals $J_R$. However, the
underlying reason for this energy dependence of $J_R$ is the smooth
variation of $J_R$ within a major shell (see above). At very small
energies one finds again a small volume integral of $J_R \approx
280$\,MeV\,fm$^3$ for $^{208}$Pb = $^{204}$Hg $\otimes$ \al . As can
also be seen from Fig.~\ref{fig:n82qlam}, the volume integrals do not
depend on the energy $E$: above $N = 82$ one finds $J_R \approx
310$\,MeV\,fm$^3$ for bound ($E < 0$) and unbound ($E > 0$) nuclei,
and below $N = 82$ one finds $J_R \approx 280$\,MeV\,fm$^3$, again for
bound and unbound nuclei.

In conclusion, systematic folding potentials can be used for the
calculation of \al -decay half-lives of superheavy
nuclei. Additionally, the systematic behavior of the volume integrals
allows to predict \al -decay energies and half-lives of yet unknown
nuclei. The magic numbers $Z_{\rm{magic}} = 132$, $N_{\rm{magic}} =
194$, and $A_{\rm{magic}} = 326$ have been derived from the
discontinuities of the volume integrals at shell closures. There is
strong evidence that \al -decay half-lives remain far below one
nanosecond even for doubly-magic superheavy nuclei above $A > 300$.

%

%
%

\end{document}